\documentclass[a4paper,12pt]{article}
\newtheorem{theorem}{Theorem}[section]

\newtheorem{remark}{Remark}[section]  
\newtheorem{assumption}{Assumption}[section]  

\usepackage{amsmath,amssymb,bm}

\title{On the Constitutive Relations in Thermo-Electroelasticity}
\author{A. Montanaro}
\author{A. Montanaro \\ montanaro$@$math.unipd.it \\
Department of Mathematics
\\University of Padua (Italy)}

\bigskip

\date{}

\begin{document}  

\maketitle



\maketitle

\centerline{\bf{Abstract}}
We give a derivation of the thermodynamic restrictions on the constitutive relations of an electrically polarizable and finitely deformable heat conducting elastic continuum, interacting with the electric field.  This is made following the method of Coleman-Noll in a thermodynamic theory with the Clausius-Duhem inequality. 

\section{Introduction}
We follow the theoretic path and some notations written in \cite{RefOncuMoodie}, where the thermodynamics restrictions for an elastic body are deduced in a theory where the method of Coleman-Noll with the Clausius-Duhem inequality are used and where, differently from here, the heat flux is an independent variable that obeys a generalized Cattaneo's equation.

This is made in order to set up the first step for extending [1] 
from
thermoelasticity to thermo-electroelasticity within a second-sound theory.
As a particular case we obtain formulae of Tiersten's fundamental paper \cite{RefTier1}, where, to describe the heat conducting continuum in interaction with the electric field,  the theory is based on a macroscopic model.
\section{Preliminary Definitions}   \label{section:2}
Let $\,{\cal E} \,$ denote a three-dimensional Euclidean point space.
We consider a body $B$ whose particles are identified with the positions 
$\,{\bm X} \in {\cal E} \,$ they occupy in a given reference configuration 
$\,{\bm B}$.   

A positive mass measure is assigned to $B$ by a referential mass density
$\, \rho_R(.) : {\bm B} \rightarrow (0, \infty)$, so that 
$\, m(P):=  \int^{}_{P}\rho_R dV\,$ is the mass of the part $P$ of ${\bm B}$.

The material filling $B$ is characterized by a given process class $\, I\!\!P(B)\,$ of $B$
as a set of ordered $10-$tuples of functions on ${\bm B} \times I\!\!R$
\begin{equation}     \label{eq:classpr}
	p= \Big( {\bm x}(.), \,  \theta(.), \,  \varphi(.), \,  \varepsilon(.), \, \eta(.) , \, {\bm \tau}(.), \,  {\bm P}(.), 
	\,  {\bm q}(.), \,  {\bm b}(.), \,  r(.) \Big) \, \in \, I\!\!P(B) 
\end{equation}
defined with respect to $\,{\bm B}\,$, satisfying the balance laws of linear momentum, moment of momentum, energy, the entropy inequality and the field equations of electrostatics,
where
\begin{itemize}
	\item $\quad {\bm x}={\bm x}({\bm X}, \, t)\,$ is the {\it motion}, 
	\item $\quad \theta=\theta({\bm X}, \, t)\in (0, \infty)\,$ is the {\it absolute temperature}, 
	\item $\quad \varphi=\varphi({\bm X}, \, t)\,$ is the {\it electric potential}, 
	\item $\quad \varepsilon=\varepsilon({\bm X}, \, t)\,$ is the specific {\it internal energy} per unit mass, 
	\item $\quad \eta=\eta({\bm X}, \, t)\,$ is the specific {\it entropy} per unit mass, 
	\item $\quad {\bm \tau}={\bm \tau}({\bm X}, \, t)\, \quad \Big({\bm S}={\bm S}({\bm X}, \, t)\Big)\,$ is the Cauchy ({\it first Piola-Kirchhoff}) stress tensor, 
	\item $\quad {\bm P}={\bm P}({\bm X}, \, t)\, \quad \Big({\bm I\!\!P}={\bm I\!\!P}({\bm X}, \, t) \Big)\,$ is the spatial (referential) {\it polarization vector}, 
	\item $\quad {\bm q}={\bm q}({\bm X}, \, t)\, \quad \Big({\bm Q}={\bm Q}({\bm X}, \, t)\Big)\,$ is the spatial (referential) {\it heat flux vector}, 
	\item $\quad {\bm b}={\bm b}({\bm X}, \, t)\,$ is the external specific {\it body force} per unit mass, 
	\item $\quad r=r({\bm X}, \, t)\,$ is the {\it radiating heating} per unit mass, 
\end{itemize}

Any motion ${\bm x}(.,.)$ of $B$ is a regular function, e.g.
two-times continuously differentiable with respect to $t$ and at each time $t$  continuous and invertible from  
$\,{\bm B}\,$ into $\,{\cal E}$. 

We use $Grad$, $grad$, $Div$, $div$, to denote gradient and divergence with respect to ${\bm X}$ and ${\bm x}$, respectively, whereas a superposed dot denotes the material time derivative.

The {\it deformation gradient} ${\bm F}$ at ${\bm X}$ at time $t$ is given by
\begin{equation}
	{\bm F}={\bm F}({\bm X}, \, t)=Grad \, {\bm x}({\bm X}, \, t) \, ,
\end{equation}
and the invertibility of the deformation is sured by the condition $$ J=det{\bm F}>0 \, .$$
 
The {\it velocity} ${\bm v}$ of ${\bm X}$ at time $t$ is given by
\begin{equation}
	{\bm v}={\bm v}({\bm X}, \, t)=\dot{\bm x}({\bm X}, \, t) \, .
\end{equation}

The law of conservation of mass is expressed by
\begin{equation}  \label{eq:cmassM}
	\rho_R = \rho J \, ,   \qquad \dot \rho + \rho div{\bm v}=0 \,,
\end{equation}
where $\;\rho=\rho({\bm X}, \, t)\;$ is the mass density of ${\bm X}$ at time $t$.

The spatial heat flux vector ${\bm q}$ and the spatial polarization vector ${\bm P}$ are related with their referential counterparts by
\begin{equation}     \label{eq:PJP}
{\bm I\!\!P}=J{\bm F}^{-1}{\bm P}\,, \qquad  {\bm Q}=J{\bm F}^{-1}{\bm q} \,.
\end{equation}

The electric potential $\varphi$, together with the polarization vector field ${\bm P}$, determines the {\it eulerian electric displacement field}, in Gaussian units, by the equality

\begin{equation}    \label{eq:euldisplvectorM}
{\bm D}={\bm E}^M+ 4\pi{\bm P} \, ,
	\end{equation}
where $\,{\bm E}^M=- \nabla_{_{\bm x}}  \varphi \,$  is the {\it (Maxwellian) spatial electric vector field}.

Note that any two corresponding referential and spatial 'energy-flux' vectors, related by
\begin{equation}    \label{eq:refdisplvectorEnFlM}
{\bm I\!\!H}=J{\bm F}^{-1}{\bm h}  \,  ,
	\end{equation}
have spatial and referential divergences that are related by
\begin{equation}    \label{eq:refdisplvectorEnFlDIVVM}
DIV {\bm I\!\!H} = J div{\bm h} \,.
	\end{equation}	
	Hence the {\it referential electric displacement field} is 
\begin{equation}    \label{eq:refdisplvectorM}
{\bm \Delta}=J{\bm F}^{-1}{\bm D}=J{\bm F}^{-1}{\bm E}^M+ 4\pi{\bm I\!\!P} \, .
	\end{equation}
	
	The {\it spatial and referential polarization vectors}  per unit mass are respectively defined by
\begin{equation}          \label{eq:eulPolarVectorUnVolM}
	{\bm \pi}={\bm P}/{\rho}  \, , \qquad {\bm \Pi}={\bm I\!\!P}/{\rho_R} \, .
\end{equation}
	\section{Spatial Description}
	\subsection{Local balance laws in spatial form}
Under suitable assumptions of regularity, and using (\ref{eq:cmassM}), the usual integral forms 
of the balance laws of linear momentum, moment of momentum, energy, the field equations of electrostatics, and the entropy inequality
are equivalent to the spatial field equations
\begin{equation}        \label{eq:eqmotS}
	\rho \dot{\bm v} = div{\bm \tau} + {\bm P} \cdot \nabla_x{\bm E}^M	+ \rho {\bm b} \, , 
\end{equation}  

\begin{equation}     \label{eq:energyS}
	\rho \dot{\varepsilon} = 
{\bm \tau} \cdot \nabla {\bm v} - div{\bm q} + {\bm E}^M \cdot \rho \dot {\bm \pi} + \rho r\, , 
\end{equation}
\begin{equation}
{\bm E}^M=- \nabla_{_{\bm x}}\varphi  \, ,   \qquad {\bm D}=\varepsilon_0 {\bm E} \,,
\end{equation} 
\begin{equation}    \label{eq:entrineqM}
	\rho \dot{\eta} \geq \rho (r/ \theta) - div({\bm q}/\theta) \, . 
\end{equation}
Incidentally, note that by the continuity equation we find
 \begin{equation}    \label{eq:PM}
{\bm E}^M \cdot \rho \dot {\bm \pi}= {\bm E}^M \cdot (\dot {\bm P} + {\bm P} div {\bm v}) \, . 
\end{equation}

Let $\,\psi=\psi()\,$ be the specific {\it free energy} per unit mass defined by
 \begin{equation}    \label{eq:freeen}
	\psi= \epsilon - \theta \eta - {\bm E}^M \cdot {\bm \pi} \, . 
\end{equation}

Then (\ref{eq:energyS}) and (\ref{eq:entrineqM}) yield the {\it dissipation inequality}
 \begin{equation}    \label{eq:DissIneq}
	\rho ( \dot \psi + \eta \dot{\theta}) - {\bm \tau}\cdot \nabla{\bm v} + \frac{1}{\theta} {\bm q} \cdot {\bm g} 
+ \rho{\bm \pi} \cdot \dot {\bm E}^M \, \leq \, 0 \, , 
\end{equation}
where $\,{\bm g}=grad\, \theta({\bm X}, \, t)\,$ is the spatial temperature gradient.

We note that Eqs. (\ref{eq:eqmotS}), (\ref{eq:energyS}), (\ref{eq:entrineqM}) and (\ref{eq:freeen}) respectively coincide with  Eqs.  (3.23), (3.40), (3.43) and (4.2) of \cite{RefTier1}.

\subsection{Constitutive Assumptions in Spatial Form}
Let $\,{\cal D}\,$ be an open, simply connected domain consisting of $\,4-$tuples
$\,( {\bm F}, \, \theta, {\bm E}^M, \, {\bm g} )$, and assume that 
if $\,( {\bm F}, \, \theta, {\bm E}^M, \, {\bm g} ) \in {\cal D}$, then
$\,( {\bm F}, \, \theta, {\bm E}^M, \, {\bm 0} ) \in {\cal D}$.

\begin{assumption} For every $\, p \in I\!\!P(B)\,$  
the specific free energy $\, \psi({\bm X}, \,t)$,
the specific entropy $\, \eta({\bm X}, \,t)$,  
the Cauchy stress tensor $\, {\bm T}({\bm X}, \,t)$,  
the specific polarization vector $\, {\bm P}({\bm X}, \,t)$, 
and the heat flux $\, {\bm q}({\bm X}, \,t)\,$ are given by continuously differentiable functions on  $\,{\cal D}\,$ such that
\begin{equation}   \label{eq:a}
	\psi= \overline{\psi}( {\bm F}, \, \theta, {\bm E}^M, \, {\bm g} )  \, ,
\end{equation}
	\begin{equation}   \label{eq:b}
	\eta= \overline{\eta}( {\bm F}, \, \theta, {\bm E}^M, \, {\bm g} )   \, ,
	\end{equation} 
	\begin{equation}   \label{eq:c}
{\bm \tau}= \overline{{\bm \tau}}( {\bm F}, \, \theta, {\bm E}^M, \, {\bm g} )  \, ,
	\end{equation}
	\begin{equation}   \label{eq:d}
{\bm P}= \overline{{\bm P}}( {\bm F}, \, \theta, {\bm E}^M, \, {\bm g} )  \, ,
	\end{equation}
	\begin{equation}   \label{eq:e}
{\bm q}= \overline{{\bm q}}( {\bm F}, \, \theta, {\bm E}^M, \, {\bm g} ) \, .
\end{equation}
\end{assumption}

Of course, once $\, \rho(.)$, $\, \overline{{\bm P}}(.)$,  $\, \overline{\psi}(.)\,$  and  $\, \overline{\eta}(.)\,$  are known, then equality (\ref{eq:freeen}) gives the continuously differentiable function $\, \overline{\varepsilon}(.)\,$ determining 
$\, \varepsilon({\bm X}, \,t) \,$ such that 
\begin{equation}   \label{eq:f}
	\varepsilon= \overline{\varepsilon}( {\bm F}, \, \theta, {\bm E}^M, \, {\bm g} ) \, .
	\end{equation}
	
Also note that the dependence upon $\,{\bm X}\,$ is not written only for brevity; when the body is not materially homogeneous it 	
becomes active.


\subsection{Coleman-Noll Method and Thermodynamic Restrictions}
Given any motion 
$\, {\bm x}({\bm X}, \, t)$, temperature field $\, \theta({\bm X}, \, t)\,$ 
 and electric potential field $\, \varphi({\bm X}, \, t)$,
the constitutive equations (\ref{eq:a})-(\ref{eq:d}) determine $\,\varepsilon({\bm X}, \, t)$, $\,\eta({\bm X}, \, t)$, $\,{\bm \tau}({\bm X}, \, t)$,
$\,{\bm P}({\bm X}, \, t)$, $\,{\bm q}({\bm X}, \, t)$, 
and the local laws (\ref{eq:eqmotS}) and   (\ref{eq:energyS}) determine $\, {\bm b}({\bm X}, \, t)\,$ 
and $\, r({\bm X}, \, t)$.
Hence for any given motion, temperature field and electric potential field a unique process $p$ is constructed.

The method of Coleman-Noll \cite{RefColNoll} is based on the postulate that every process $p$ so constructed belongs to the process class $\,I\!\!P(B)\,$ of $B$, that is, on the assumption that the constitutive assumptions (\ref{eq:a})-(\ref{eq:e}) are compatible with thermodynamics, in the sense of the following

\bigskip 

{\bf Dissipation Principle} {\it $\;$ 
For any given motion, temperature field and electric potential field, the process $p$ constructed from  the constitutive equations (\ref{eq:a})-(\ref{eq:e}) belongs to the process  class $\,I\!\!P(B)\,$ of $B$.
Therefore the constitutive functions (\ref{eq:a})-(\ref{eq:e}) are compatible with the second law of thermodynamics in the sense that they satisfy the dissipation inequality
(\ref{eq:DissIneq}). }

	\medskip
	
It is a matter of routine to extend to thermo-electroelasticity the remark for thermoelasticity written in \cite{RefColNoll},  on page 1119, lines 8-30 from top; 
here we write such extension by paraphrasing Coleman's remark.
\begin{remark}          \label{remark:ColMizRem}
Let ${\bm A}(t)$ be any time-dependent invertible tensor, 
$\alpha(t)$ any time-dependent positive scalar,  
${\bm a}(t)$ any time-dependent vector, 
$\beta(t)$ be any time-dependent scalar, ${\bm b}(t)$ any time-dependent vector, and 
$\,Y\,$ any material point of $\,B \,$ whose spatial position in the reference configuration $\,{\bm B}\,$ is ${\bm Y}$.
We can always construct at least one admissible electro-thermodynamic process in 
$\,{\bm B}\,$ such that  
$${\bm F}({\bm X}, \,t ) , \; \theta({\bm X}, \,t ), \; {\bm g}({\bm X}, \,t ),\; {\bm E}^M({\bm X}, \,t ) $$ 
have, respectively, the values
$\,{\bm A}(t), \, \alpha(t),\, {\bm a}, \, {\bm b}\,$ 
at $\, {\bm X}={\bm Y}$.

An example of such a process is the one determined by the following deformatoin function, temperature distribution and electric potential:
\begin{equation}
	{\bm x}={\bm x}({\bm X}, \, t)={\bm Y}+{\bm A}(t)[{\bm X}-{\bm Y}]\,,
\end{equation}
\begin{equation}
	\theta=\theta({\bm X}, \, t)=\alpha(t)+[{\bm A}^T(t){\bm a}(t)] \cdot [{\bm X}-{\bm Y}]\,,
\end{equation}
\begin{equation}
	\varphi=\varphi({\bm X}, \, t)=\beta(t)+[{\bm A}^T(t){\bm b}(t)] \cdot [{\bm X}-{\bm Y}]\,.
\end{equation}
Thus, at a given time  $t$, we can arbitrarily specify not only 
$\,{\bm F}, \; \theta, \; {\bm g}\, $ and  $\,{\bm E}^M \,$
but also their time derivatives
$\,\dot{\bm F} , \; \dot\theta, \; \dot{\bm g}\, $ and  $\,\dot{\bm E}^M\,$ at a point $\,{\bm Y}\,$ and be sure that there exists at least one electro-thermodynamic process corresponding to this choice. 
\end{remark}
 
The next theorem can be proved by using this remark.

\begin{theorem}   \label{theorem:dissPrSpatial}
The Dissipation Principle is satisfied if and only if the following conditions hold:

(i) the free energy response function 
$\,\overline{\psi}({\bm F}, \, \theta, {\bm E}^M, \,  {\bm g} )  \,$
is independent of the temperature gradient ${\bm g}$ and determines the entropy,  the first Piola-Kirchhoff stress, and the polarization vector through the relations
\begin{equation}   \label{eq:constRestreta}
\overline{\eta}({\bm F}, \, \theta, {\bm E}^M)=
	-\partial_{\theta} \overline{\psi}({\bm F}, \, \theta, {\bm E}^M)  \, , 
\end{equation}

\begin{equation}   \label{eq:constRestrK}
\overline{{\bm \tau}}({\bm F}, \, \theta, {\bm E}^M)=
\rho {\bm F} \partial_{{\bm F}} \overline{\psi}({\bm F}, \, \theta, {\bm E}^M)
\, , 
\end{equation}

\begin{equation}   \label{eq:constRestrP}
\overline{{\bm \pi}}({\bm F}, \, \theta, {\bm E}^M)
 =-\partial_{{\bm E}^M} \overline{\psi}({\bm F}, \, \theta, {\bm E}^M)\, .
\end{equation}

(ii) the reduced dissipation inequality (Fourier inequality)
\begin{equation}    \label{eq:RedDissIneq}
{\bm q} \cdot {\bm g} \, \leq \, 0 \, 
\end{equation}
is satisfied.  \end{theorem}
\smallskip

\underline{Proof}.  By the chain rule we have
\begin{equation}   \label{eq:psiChainRule}
	\dot \psi = \partial_{{\bm F}} \overline{\psi} \cdot \dot {\bm F}
	            + \partial_{\theta} \overline{\psi} \cdot \dot \theta
	            + \partial_{{\bm E}^M} \overline{\psi} \cdot \dot {\bm E}
	            + \partial_{{\bm g}} \overline{\psi} \cdot \dot {\bm g}  \, .
\end{equation}

Note that
\begin{equation}  \label{eq:gradvFF}
	   \frac{\partial v^i}{\partial x^j}=  \frac{\partial v^i}{\partial X^K} \frac{\partial X^K}{\partial x^j}
	   =\frac{\partial X^K}{\partial x^j}\frac{d}{dt}\frac{\partial x^i}{\partial X^K} \, , \end{equation}  
that is,
\begin{equation}  \label{eq:gradvFFAbs}
	   \nabla \bf v = {\bm F}^{-T} \dot {\bm F} \, , \end{equation}  
	   and thus
\begin{equation}  \label{eq:gradvFFAbs++}
{\bm \tau} \cdot  \nabla \bf v = {\bm F}^{-1} {\bm \tau} \cdot \dot{\bm F} \, . \end{equation}  

Thus by substituting Eqn. (\ref{eq:psiChainRule}) together with the constitutive equations (\ref{eq:a})-(\ref{eq:e}) into the dissipation inequality (\ref{eq:DissIneq}), gives

\begin{eqnarray}                               \label{eq:ConsPsiChainRule}
	(\rho \partial_{{\bm F}} \overline{\psi} -{\bm F}^{-1} \overline{{\bm \tau}}) \cdot \dot {\bm F}
	+ 	(\rho \partial_{\theta} \overline{\psi} + \overline{\eta}) \dot \theta
	+ 	(\rho \partial_{{\bm E}^M} \overline{\psi} + \overline{{\bm P}}) \cdot \dot {\bm E}^M 
	 \qquad \qquad \\
	+ 	\rho \partial_{{\bm g}} \overline{\psi} \cdot \dot {\bm g} 
	+ \frac{1}{\theta} {\bm q} \cdot {\bm g} 
  \, \leq \, 0 \, .
\end{eqnarray}

Now we proceed in the Coleman-Mizel \cite{RefColNoll} method: by Remark \ref{remark:ColMizRem} we can we can state that 
$\,\dot {\bm F}, \, \dot \theta, \, \dot {\bm E}^M\,$ and  $\, \dot {\bm g} \,$
can be assigned arbitrary values independently from the other variables and 
the theorem is proved. $\; \diamondsuit$

\medskip
A consequence of the reduced dissipation inequality (\ref{eq:RedDissIneq}) is that, just as in thermoelasticity, the {\it static heat flux} vanishes:
\begin{theorem}
The heat flux $\, {\bm q}\,$ vanishes for all termal equilibrium states 
$\,( {\bm F}, \, \theta, {\bm E}^M, \, {\bm 0} ) \in {\cal D}$, that is, 
\begin{equation}
\overline{{\bm q}}( {\bm F}, \, \theta, {\bm E}^M, \, {\bm 0} )
\,=\, {\bm 0}  \, .
\end{equation}
\end{theorem}
\subsection{Use of Invariant response functions}
In order to satisfy the principle of material objectivity  \cite{RefToupin} , \cite{RefEringen}, the functions $\; \varepsilon$ and $\psi$, must be a scalar invariant under rigid rotations of the deformed and polarized body.
The invariance of  $\psi$ in a rigid rotation is assured when  $\psi$ is an arbitrary function of the referential quantities $\,E_{LM},\, \theta, \, W_{L}, \, G_{L}$,
where 
\begin{equation}   \label{eq:ERS}
	E_{LM}=\frac{1}{2}(C_{LM}-\delta_{LM}), \qquad   C_{LM}=x_{k,\,L}x_{k,\,M} \, ,
\end{equation}

\begin{equation}  \label{eq:WE}
 W_L=-\frac{\partial \varphi}{\partial X_L} = -\frac{\partial \varphi}{\partial x^p} \frac{\partial x^p}{\partial X_L}\,, \qquad  {\bm W}={\bm F}^{T}{\bm E}^M \, ,
\end{equation}
\begin{equation}   \label{eq:QqGg}
\frac{\partial G}{\partial X_L} = \frac{\partial G}{\partial x^p} \frac{\partial x^p}{\partial X_L}\,,  \qquad
{\bm G}={\bm F}^{T}{\bm g} \, .
\end{equation}

Hence we assume that 
\begin{equation}     \label{eq:psi=hatpsi}
	\psi=\tilde \psi( {\bm E}, \, \theta, {\bm W}, \, {\bm G} )  \, .
\end{equation}

Next we calculate the time derivatives in equation (\ref{eq:psiChainRule}) by using $\tilde\psi$ in place of $\overline{\psi}$. 
We find 
\begin{equation} \label{eq:psitime1}
	 \frac{\partial \overline{\psi}}{\partial \bf F}\cdot \dot{\bm F} 
	 = \Big[ \frac{\partial \tilde \psi}{\partial E_{RS}} \, \frac{\partial E_{RS}}{\partial (\partial x^i/\partial X_K)}  
	 + \frac{\partial \tilde \psi}{\partial W_R}  \frac{\partial W_R}{\partial (\partial x^i/\partial X_K)} 
	 + \frac{\partial \tilde \psi}{\partial G_R}  \frac{\partial G_R}{\partial (\partial x^i/\partial X_K)} 
	 \Big] \frac{d}{dt}\frac{\partial x^i}{\partial X_K} \\
\end{equation}

Now, by (\ref{eq:ERS})-(\ref{eq:QqGg}),

\begin{eqnarray} \label{eqnarray:psitime11}  \nonumber
   \frac{\partial \tilde \psi}{\partial E_{RS}} \, \frac{\partial E_{RS}}{\partial (\partial x^i/\partial X_K)}   \frac{d}{dt}\frac{\partial x^i}{\partial X_K}
=
\frac{\partial \tilde \psi}{\partial E_{RS}}
 \frac{1}{2}\Big( \delta_{RK}\frac{\partial x^i}{\partial X_S} + \frac{\partial x^i}{\partial X_R}\delta_{SK}
  \Big) \frac{\partial \dot x^i}{\partial X_K} \\
= \frac{1}{2} \Big( 
   \frac{\partial \tilde \psi}{\partial E_{KS}}\frac{\partial x^i}{\partial X_S} 
   + \frac{\partial \tilde \psi}{\partial E_{RK}}\frac{\partial x^i}{\partial X_R} 
  \Big) \frac{\partial \dot x^i}{\partial X_K}
  =  \frac{\partial \tilde \psi}{\partial E_{RK}}\frac{\partial x^i}{\partial X_R} 
  \frac{\partial \dot x^i}{\partial X_K} = \Big( \frac{\partial \tilde \psi}{\partial {\bm E}}{\bm F}^T \Big) \cdot \dot {\bm F}  \, ,
\end{eqnarray}

\begin{equation}
	\frac{\partial \tilde \psi}{\partial W_R}  \frac{\partial W_R}{\partial (\partial x^i/\partial X_K)}\frac{d}{dt}\frac{\partial x^i}{\partial X_K} =
	\frac{\partial \tilde \psi}{\partial W_R}  \delta_{KR} E^M_i\frac{d}{dt}\frac{\partial x^i}{\partial X_K}=
	\Big( \frac{\partial \tilde \psi}{\partial {\bm W}} \otimes
 {\bm E}^M \Big)\cdot \dot {\bm F} \, ,
\end{equation}
and, similarly with the latter,
\begin{equation}
	\frac{\partial \tilde \psi}{\partial G_R}  \frac{\partial G_R}{\partial (\partial x^i/\partial X_K)}\frac{d}{dt}\frac{\partial x^i}{\partial X_K} =
	\frac{\partial \tilde \psi}{\partial G_R}  \delta_{KR} g_i\frac{d}{dt}\frac{\partial x^i}{\partial X_K}=
	\Big( \frac{\partial \tilde \psi}{\partial {\bm G}} \otimes
 {\bm g} \Big)\cdot \dot {\bm F} \, ;
\end{equation}
	hence 
	
\begin{equation} \label{eq:psitime1111}
	 \frac{\partial \overline{\psi}}{\partial \bf F} \cdot \dot {\bm F}
	 = \Big( \frac{\partial \tilde \psi}{\partial {\bm E}}{\bm F}^T 
	 +  \frac{\partial \tilde \psi}{\partial {\bm W}} \otimes
 {\bm E}^M 
	 + \frac{\partial \tilde \psi}{\partial {\bm G}} \otimes
 {\bm g} \Big)\cdot \dot {\bm F} \, ;
\end{equation}

moreover,
\begin{equation} \label{eq:psitime223}
	 \frac{\partial \overline{\psi}}{\partial \bf g}\cdot \dot{\bm g} 
=
 \frac{\partial \tilde \psi}{\partial G_R}  \frac{\partial G_R}{\partial g^i} \dot g^i
 =   \frac{\partial \tilde \psi}{\partial G_R}  F^i_R \dot g^i 
=\Big({\bm F}\frac{\partial \tilde \psi}{\partial \bf G}\Big)\cdot \dot{\bm g} \, ,
\end{equation}
	
\begin{equation} \label{eq:psitime1122}  \nonumber
\frac{\partial \overline{\psi}}{\partial {\bm E}^M}\cdot \dot {\bm E}^M=
 \Big({\bm F} \frac{\partial \tilde \psi}{\partial {\bm W}}\Big) \cdot \dot {\bm E}^M  \, .
\end{equation}
	
By recollecting the equalities above we can rewrite the {\it dissipation inequality} (\ref{eq:DissIneq}) as
 \begin{eqnarray}    \label{eqnarray:DissIneq2} \nonumber
	\rho \Big[ 
	\Big(\frac{\partial \tilde \psi}{\partial {\bm E}}{\bm F}^T 
	 +  \frac{\partial \tilde \psi}{\partial {\bm W}} \otimes
 {\bm E}^M 
	 + \frac{\partial \tilde \psi}{\partial {\bm G}} \otimes
 {\bm g} \Big)\cdot \dot {\bm F}  
	            + \frac{\partial \tilde \psi}{\partial \theta} \cdot \dot \theta
	            + \Big({\bm F} \frac{\partial \tilde \psi}{\partial {\bm W}}\Big) \cdot \dot {\bm E}^M  
	            + \Big({\bm F}\frac{\partial \tilde \psi}{\partial \bf G}\Big)\cdot \dot{\bm g}
	           + \eta \dot{\theta}\Big] \\ 
	            - 	{\bm \tau} \cdot \nabla{\bm v} + \frac{1}{\theta} {\bm q} \cdot {\bm g} 
+ \rho{\bm \pi} \cdot \dot {\bm E}^M \, \leq \, 0 \, .   \qquad \qquad \qquad \qquad
\end{eqnarray}

Now we apply the method of Colemann-Noll in Remark \ref{remark:ColMizRem}.
By the arbitrariness of $\,{\bm g}\,$ we have $\,\partial \tilde \psi/{\partial \bf G}=\bf O$.
Remembering (\ref{eq:gradvFFAbs}), 
(\ref{eq:gradvFFAbs++})
	   Eq. (\ref{eqnarray:DissIneq2}) reduces to
	   
	   \begin{eqnarray}    \label{eqnarray:DissIneq23} \nonumber
	\rho \Big[ 
	\Big( \frac{\partial \tilde \psi}{\partial {\bm E}} {\bm F}^T
	 +  \frac{\partial \tilde \psi}{\partial {\bm W}} \otimes
 {\bm E}^M - \rho^{-1}{\bm F}^{-1} {\bm \tau} \Big)\cdot \dot {\bm F}  
 + \Big( \frac{\partial \tilde \psi}{\partial \theta}+\eta \Big) \cdot \dot \theta \\  
	            + \Big({\bm F} \frac{\partial \tilde \psi}{\partial {\bm W}}
	            + {\bm \pi} \Big) \cdot \dot {\bm E}^M \Big] 
	           + \frac{1}{\theta} {\bm q} \cdot {\bm g} \, \leq \, 0 \, .   
\end{eqnarray}

By the arbitrariness of the time derivatives $\, \dot {\bm F}$,$\, \dot \theta$, 
$\, \dot {\bm E}^M\,$ and by substituting the constitutive relation (\ref{eq:e}) we find
\begin{equation}   \label{eq:TPsi}
	\rho^{-1}{\bm F}^{-1} {\bm \tau} = \frac{\partial \tilde \psi}{\partial {\bm E}} {\bm F}^T
	 +  \frac{\partial \tilde \psi}{\partial {\bm W}} \otimes {\bm E}^M 
\end{equation}
\begin{equation}   \label{eq:etaPsi}
	\eta = - \frac{\partial \tilde \psi}{\partial \theta} \, ,
\end{equation}
\begin{equation}   \label{eq:piPsi}
{\bm \pi}=-	{\bm F} \frac{\partial \tilde \psi}{\partial {\bm W}} \, ,
\end{equation}
	   \begin{equation}    \label{eq:DissIneq23} \nonumber
 \frac{1}{\theta} \, {\bm q} \cdot {\bm g} \, \leq \, 0 \, .   
\end{equation}

We have proved the version of Theorem \ref{theorem:dissPrSpatial} that employes the objective energy response function $\, \tilde \psi$.

\begin{theorem}   \label{theorem:dissPrSpatialHAT}
The Dissipation Principle is satisfied if and only if the following conditions hold:

(i) the objective free energy response function 
$\,\tilde \psi( {\bm E}, \, \theta, {\bm W},\, {\bm G} )  \,$
is independent of the temperature gradient ${\bm G}$ and determines the entropy,  the Cauchy stress tensor, and the polarization vector per unit mass through the relations (\ref{eq:TPsi})-(\ref{eq:piPsi});

(ii) the reduced dissipation inequality (\ref{eq:DissIneq23})
is satisfied.  
  \end{theorem}

We point out that Equalities (\ref{eq:TPsi}), (\ref{eq:eulPolarVectorUnVolM}) and (\ref{eq:piPsi}) yield the following expression for the Cauchy stress: 
\begin{equation}   \label{eq:TPsiExpr}
	{\bm \tau} = \rho {\bm F} \frac{\partial \tilde \psi}{\partial {\bm E}} {\bm F}^T
	 -  {\bm P} \otimes {\bm E}^M \, ;
\end{equation}
hence for the antisymmetric portion $\,{\bm \tau}^A\,$  of $\,{\bm \tau}\,$ we obtain the expression
\begin{equation}   \label{eq:TPsiExprSKW}
{\bm \tau}^A = \frac{1}{2} \,\Big( {\bm E}^M \otimes {\bm P} -  {\bm P} \otimes {\bm E}^M \Big) \, ,
\end{equation}
that coincides with (3.24) of \cite{RefTier1}.


\subsection{Internal Dissipation and Entropy Equality}   
The local {\it internal dissipation} $\,\delta_o\,$ in a thermoelastic body is defined by \cite{RefTruesdell} (p.112)
\begin{equation}\label{eq:IntDissip}
\delta_o=\theta \dot \eta -(r -\frac{1}{\rho}div{\bm q})
\end{equation}
Then it is proved that $\, \delta_o \equiv 0 \,$ along every local thermoelastic process. Here, with regard to thermo-electroelasticity, we define the {\it internal dissipation} just by (\ref{eq:IntDissip}) and hence we extend the above theorem by the following
\begin{theorem}\label{theorem:diss}
Along any local process of $B$ we have
\begin{equation}\label{eq:Thdiss}
	\delta_o= 0   \, .
\end{equation}
\end{theorem}
\underline{Proof}.  

By inserting the energy equation (\ref{eq:energyS}) in the definition (\ref{eq:IntDissip}) we obtain the equality 
\begin{equation}\label{eq:disssec}
	\delta_o=\theta \dot \eta 
-\frac{1}{\rho}\big(\rho \dot \varepsilon - {\bm \tau}\cdot \nabla {\bm v} 
          - {\bm E}^M \cdot \rho \dot{{\bm \pi}} \big)   \end{equation}

          Now by (\ref{eq:freeen}) we find 
          
\begin{equation}
	\theta \dot \eta 
	= - \dot \psi + \dot \varepsilon - \dot \theta \eta 
	- \dot {\bm E}^M \cdot {\bm \pi} 
	- {\bm E}^M \cdot \dot{{\bm \pi}} 
\end{equation}
and by replacing the latter into (\ref{eq:disssec})  we have

\begin{equation}\label{eq:disssec2}
	\delta_o=
	\Big(- \dot \psi + \dot \varepsilon - \dot \theta \eta 
	- \dot {\bm E}^M \cdot {\bm \pi} 
	- {\bm E}^M \cdot \dot{{\bm \pi}} \Big) 
-\frac{1}{\rho}\Big(\rho \dot \varepsilon - {\bm \tau}\cdot \nabla {\bm v} 
          - {\bm E}^M \cdot \rho \dot{{\bm \pi}} \Big)  \, , \end{equation}
thus
\begin{equation}\label{eq:disssec3}
	\delta_o=
	\Big(- \dot \psi - \dot \theta \eta 
	- \dot {\bm E}^M \cdot {\bm \pi} \Big) 
+ \frac{1}{\rho} \, {\bm \tau}\cdot \nabla {\bm v}  \, . \end{equation}

Now by (\ref{eq:psiChainRule})$_1$
\begin{equation}\label{eq:disssec4}
	\delta_o=
	\Big(- \partial_{{\bm F}} \overline{\psi} \cdot \dot {\bm F}
	           - \partial_{\theta} \overline{\psi} \cdot \dot \theta
	           -\partial_{{\bm E}^M} \overline{\psi} \cdot \dot {\bm E}^M
	            -\partial_{{\bm g}} \overline{\psi} \cdot \dot {\bm g}  \,
-\dot \theta \eta 
	- \dot {\bm E}^M \cdot {\bm \pi} \Big) 
+ \frac{1}{\rho} \, {\bm \tau}\cdot \nabla {\bm v}  \, \end{equation}
and the constitutive restrictions 
(\ref{eq:constRestreta})-(\ref{eq:constRestrP}) together with (\ref{eq:gradvFFAbs++}) yield (\ref{eq:Thdiss}).
$\; \diamondsuit$

\bigskip 

In a thermoelastic body any thermoelastic process is locally reversible, in the sense that the following entropy equality holds

\begin{equation}  \label{eq:EntEq}
	\rho \dot \eta = \rho \frac{r}{\theta} - \frac{div{\bm q}}{\theta} \, .
\end{equation}

The last theorem and (\ref{eq:IntDissip}) yield the same result in thermo-electroelasticity too.
\begin{theorem}\label{theorem:LocRev}
Along any local process of $B$ the entropy equality
(\ref{eq:EntEq}) holds.
\end{theorem}

\section{Referential description}

We can rewrite the constitutive relations (\ref{eq:a})-(\ref{eq:e}) in material form by using the first Piola-Kirchhoff stress tensor $\, {\bm S}({\bm X}, \,t)$, that is related with the Cauchy stress by

\begin{equation}
	{\bm S}=J{\bm F}^{-1}{\bm \tau} \, 
\end{equation}
and by using the well known equalities (\ref{eq:PJP}), (\ref{eq:refdisplvectorM})-(\ref{eq:refdisplvectorEnFlDIVVM}) and
\begin{equation}
	Div{\bm S}=Jdiv{\bm \tau} \, ,   
\end{equation}
Now the process class $\, I\!\!P(B)\,$ of $B$ of Section \ref{section:2}, contatining the processes (\ref{eq:classpr}), must be substituted with  $\,I\!\!P_R(B)$, that is the 
set of ordered $10-$tuples of functions on ${\bm B} \times I\!\!R$
\begin{equation}
	p_R= \Big( {\bm x}(.), \,  \theta(.), \,  \varphi(.), \,  \varepsilon(.), \, \eta(.) , \,  {\bm S}(.), \,  {\bm I\!\!P}(.), 
	\,  {\bm Q}(.), \,  {\bm b}(.), \,  r(.) \Big) \, \in \, I\!\!P_R(B) 
\end{equation}
defined with respect to $\,{\bm B}\,$, satisfying the material versions of the balance laws of linear momentum, moment of momentum, energy, the entropy inequality,and the field equations of electrostatics.

	\subsection{Local Balance Laws in Material Form}
Under suitable assumptions of regularity, and using (\ref{eq:cmassM}), the usual integral forms 
of the balance laws of linear momentum, moment of momentum, energy, the field equations of electrostatics, and the entropy inequality
are equivalent to the referential field equations
 \begin{equation}        \label{eq:eqmot}
	\rho_R \dot{\bm v} = Div {\bm S} + 	\rho_R {\bm b} \, , 
\end{equation}  

 \begin{equation}     \label{eq:energy}
	\rho_R \dot{\varepsilon} = 
	{\bm S} \cdot \dot{\bm F} - Div{\bm Q} + {\bm W} \cdot \dot {\bm I\!\!P} + \rho_R r\, , 
\end{equation}

  \begin{equation}
{\bm W}=- \nabla_{_{\bm X}}\varphi \;(=-{\bm F}^T \nabla_{_{\bm x}}\varphi )  \,  , \qquad Div {\bm \Delta} = 0 \, , 
\end{equation}

 \begin{equation}    \label{eq:entrineq}
	\rho_R \dot{\eta} \geq \rho_R (r/ \theta) - Div({\bm Q}/\theta) \, . 
\end{equation}

Let $\,\psi=\psi()\,$ be the specific {\it free energy} per unit mass defined by
 \begin{equation}    \label{eq:freeenS}
	\psi= \varepsilon - \theta \eta -  {\bm W} \cdot {\bm \Pi} \, . 
\end{equation}

Then (\ref{eq:energy}) and (\ref{eq:entrineq}) yield the {\it dissipation inequality}
 \begin{equation}    \label{eq:DissIneqM}
	\rho_R ( \dot \psi + \eta \dot{\theta}) - 	{\bm S} \cdot \dot{\bm F} +  \frac{1}{\theta} {\bm Q} \cdot {\bm G} 
+ {\bm I\!\!P} \cdot \dot {\bm W} \, \leq \, 0 \, , 
\end{equation}
where $\,{\bm G}=Grad\, \theta({\bm X}, \, t)\,$ is the referential temperature gradient.

\begin{remark}
Note that the free-energy function $\psi$ defined here by (\ref{eq:freeenS}) coincides with the analogous function $\chi$ defined in (4.2) of 
\cite{RefTier1}, on page 596.
In fact, 
$${\bm W} \cdot {\bm \Pi}= {\bm W} \cdot {\bm I\!\!P}/\rho_R 
= ({\bm F}^T {\bm E}^M) \cdot (J{\bm F}^{-1}{\bm P})/\rho_R
={\bm E}^M \cdot {\bm P} (J/\rho_R) \, $$ 
Hence, by $\,\rho_R=J\rho$, we have
$${\bm W} \cdot {\bm \Pi}
={\bm E}^M \cdot {\bm P}/\rho
= {\bm E}^M \cdot {\bm \pi}\, ,$$
where $\, \pi\,$ is the spatial polarization vector. 
\end{remark}

\subsection{Referential Constitutive Assumptions}
Let $\,{\cal D}_R\,$ be the open, simply connected domain consisting of $\,4-$tuples
$\,( {\bm F}, \, \theta, \,{\bm W}, \, {\bm G} )\,$  such that
$\,( {\bm F}, \, \theta, \,{\bm W},\, {\bm G} ) \in {\cal D}$;
hence if $\,( {\bm F}, \, \theta, \, {\bm W},\, {\bm G} ) \in {\cal D}_R$, then
$\,( {\bm F}, \, \theta, \, {\bm W}, \, {\bm 0} )  \in  {\cal D}_R\,$.

Next we use a free energy function of the form
\begin{equation}     \label{eq:psi=overpsi}
	\psi=\hat{\psi}( {\bm F}, \, \theta, \, {\bm W},\, {\bm G} )  \, .
\end{equation}
\begin{assumption} For every $\, p \in I\!\!P_R(B)\,$  
the specific free energy $\, \psi({\bm X}, \,t)$,
the specific entropy $\, \eta({\bm X}, \,t)$,  
the first Piola-Kirchhoff stress tensor $\, {\bm S}({\bm X}, \,t)$,  
the specific polarization vector $\, {\bm I\!\!P}({\bm X}, \,t)$, 
and the heat flux $\, {\bm Q}({\bm X}, \,t)\,$ are given by continuously differentiable functions on  $\,{\cal D}_R\,$ such that
\begin{equation}   \label{eq:aM}
	\psi= \hat{\psi}( {\bm F}, \, \theta, {\bm W},\, {\bm G} )  \, ,
\end{equation}
	\begin{equation}   \label{eq:bM}
	\eta= \hat{\eta}( {\bm F}, \, \theta, {\bm W},\, {\bm G} )   \, ,
	\end{equation} 
	\begin{equation}   \label{eq:cM}
	{\bm S}= \hat{{\bm S}}( {\bm F}, \, \theta, {\bm W},\, {\bm G} )  \, ,
	\end{equation}
	\begin{equation}   \label{eq:dM}
{\bm I\!\!P}= \hat{{\bm I\!\!P}}( {\bm F}, \, \theta, {\bm W},\, {\bm G} )  \, ,
	\end{equation}
	\begin{equation}   \label{eq:eM}
{\bm Q}= \hat{\bm Q}({\bm F}, \, \theta, {\bm W},\, {\bm G} ) \, .
\end{equation}
\end{assumption}

Of course, once $\, \rho_R(.)$, $\, \hat {\bm I\!\!P}(.)$,  $\, \hat \psi(.)\,$  and  $\, \hat \eta(.)\,$  are known, then equality (\ref{eq:freeenS}) gives the continuously differentiable function $\, \hat \varepsilon(.)\,$ determining 
$\, \varepsilon({\bm X}, \,t) \,$ such that 
\begin{equation}   \label{eq:fM}
	\varepsilon= \hat \varepsilon( {\bm F}, \, \theta, {\bm W}, \, {\bm G} )  \, .
	\end{equation}
	
Also note that the dependence upon $\,{\bm X}\,$ is not written for convenience, but it is implicit and understood when the body is not materially homogeneous.
\subsection{Coleman-Noll Method and Thermodynamic Restrictions}
Given any motion 
$\, {\bm x}({\bm X}, \, t)$, temperature field $\, \theta({\bm X}, \, t)\,$ 
 and electric potential field $\, \varphi({\bm X}, \, t)$,
the constitutive equations (\ref{eq:aM})-(\ref{eq:dM}) determine $\,e({\bm X}, \, t)$, 
$\,\eta({\bm X}, \, t)$, $\,{\bm S}({\bm X}, \, t)$,
$\,{\bm I\!\!P}({\bm X}, \, t)$, 
and the local laws (\ref{eq:eqmot}) and   (\ref{eq:energy}) determine $\, {\bm b}({\bm X}, \, t)\,$ 
and $\, r({\bm X}, \, t)$.
Hence for any given motion, temperature field and electric potential field, a unique process $p$ is constructed.

The method of Coleman-Noll \cite{RefColNoll} is based on the postulate that every process $p$ so constructed belongs to the process class $\,I\!\!P(B)\,$ of $B$, that is, on the assumption that the constitutive assumptions (\ref{eq:aM})-(\ref{eq:eM}) are compatible with thermodynamics, in the sense of the following

\bigskip 

{\bf Dissipation Principle} {\it $\;$ 
For any given motion, temperature field and electric potential field, the process $p$ constructed from  the constitutive equations (\ref{eq:aM})-(\ref{eq:eM}) belongs to the process  class $\,I\!\!P(B)\,$ of $B$.
Therefore the constitutive functions (\ref{eq:aM})-(\ref{eq:eM}) are compatible with the second law of thermodynamics in the sense that they satisfy the dissipation inequality
(\ref{eq:entrineq}). }

\begin{theorem}
The Dissipation Principle is satisfied if and only if the following conditions hold:

(i) the free energy response function 
$\,\hat \psi( {\bm F}, \, \theta, {\bm W},\, {\bm G} )  \,$
is independent of the temperature gradient ${\bm G}$ and determines the entropy,  the first Piola-Kirchhoff stress, and the polarization vector per unit mass \eqref{eq:eulPolarVectorUnVolM}$_2$ through the relations
\begin{equation}   \label{eq:constRestretaM}
\hat\eta({\bm F}, \, \theta, {\bm W})=-\partial_{\theta} \hat \psi({\bm F}, \, \theta, {\bm W})  \, , 
\end{equation}
\begin{equation}   \label{eq:constRestrKM}
\hat{\bm S}({\bm F}, \, \theta, {\bm W}, \, {\bm Q})=
\rho_R \partial_{{\bm F}} \hat \psi({\bm F}, \, \theta, {\bm W}, \, {\bm Q})
\end{equation}
\begin{equation}   \label{eq:constRestrPM}
\hat{\bm \Pi}({\bm F}, \, \theta, {\bm W}, \, {\bm Q})
 =-\partial_{{\bm W}} \hat \psi({\bm F}, \, \theta, {\bm W}, \, {\bm Q})
\end{equation}

(ii) the reduced dissipation inequality
\begin{equation}    \label{eq:RedDissIneqM}
{\bm Q} \cdot {\bm G} \, \leq \, 0 \, 
\end{equation}
is satisfied.  \end{theorem}
\smallskip

\underline{Proof}.  By he chain rule we have
\begin{equation}   \label{eq:psiChainRuleM}
	\dot \psi = \partial_{{\bm F}} \hat \psi \cdot \dot {\bm F}
	            + \partial_{\theta} \hat \psi \cdot \dot \theta
	            + \partial_{{\bm W}} \hat \psi \cdot \dot {\bm W}
	            + \partial_{{\bm G}} \hat \psi \cdot \dot {\bm G}  \, .
\end{equation}

Thus by substituting this equation together with the constitutive equations (\ref{eq:aM})-(\ref{eq:eM}) into the dissipation inequality (\ref{eq:DissIneqM}) gives

\begin{eqnarray}                               \label{eq:ConsPsiChainRuleM}
	(\rho_R \partial_{{\bm F}} \hat \psi - \hat {\bm S}) \cdot \dot {\bm F}
	+ 	(\rho_R \partial_{\theta} \hat \psi + \hat \eta) \dot \theta
	+ 	(\rho_R \partial_{{\bm W}} \hat \psi + \hat {\bm I\!\!P}) \cdot \dot {\bm W} 
	 \qquad \qquad \\
	+ 	\rho_R \partial_{{\bm G}} \hat \psi \cdot \dot {\bm G} 
	+ \frac{1}{\theta} {\bm Q} \cdot {\bm G} 
  \, \leq \, 0 \, .
\end{eqnarray}

Now we use Coleman-Mizel method \cite{RefColNoll}: by Remark \ref{remark:ColMizRem}, translated in referential form, we have that 
$\,\dot {\bm F}, \, \dot \theta, \, \dot {\bm W}\,$ and  $\, \dot {\bm G}\,$
can be assigned arbitrary values independently from the other variables and the theorem is easily proved just as in the proof of Theorem \ref{theorem:dissPrSpatial}. 
$\; \diamondsuit$

\bigskip

A consequence of the reduced dissipation inequality (\ref{eq:RedDissIneqM}) is that, just as in thermoelasticity, the {\it static heat flux} vanishes:
\begin{theorem}
The heat flux $\, {\bm Q}\,$ vanishes for all termal equilibrium states 
$\,( {\bm F}, \, \theta, {\bm E}^M, \, {\bm 0} ) \in {\cal D}$, that is, 
\begin{equation}
\hat{{\bm Q}}( {\bm F}, \, \theta, {\bm W}, \, {\bm 0} )
\,=\, {\bm 0}  \, .
\end{equation}
\end{theorem}


\end{document}